\documentclass[sigconf, nonacm]{acmart}
\usepackage{soul}
\usepackage{subfigure}
\usepackage{amsmath}
\usepackage{xcolor}
\AtBeginDocument{%
  \providecommand\BibTeX{{%
    \normalfont B\kern-0.5em{\scshape i\kern-0.25em b}\kern-0.8em\TeX}}}

\begin{document}


\title{Axon: A novel systolic array architecture for improved run time and energy efficient GeMM and Conv operation  with on-chip im2col}
\author{Md Mizanur Rahaman Nayan,  Ritik Raj, Gouse Basha Shaik, Tushar Krishna, 
Azad J Naeemi \\ Department of Electrical and Computer Engineering \\ Georgia Institute of Technology, USA}

\begin{abstract}
General matrix multiplication (GeMM) is a core operation in virtually all AI applications. Systolic array (SA) based architectures have shown great promise as GeMM hardware accelerators thanks to their speed and energy efficiency.
Unfortunately, SAs incur a linear delay in filling the operands, due to unidirectional propogation via pipeline latches. 
In this work, we propose a novel in-array data orchestration technique in SAs where we enable data feeding on the principal diagonal followed by bi-directional propagation. This improves the runtime by up to $2\times$ at minimal hardware overhead.
In addition, the proposed data orchestration enables convolution lowering (known as im2col) using a simple hardware support to fully exploit input feature map reuse opportunity and significantly lower the off-chip memory traffic resulting in $1.2\times$ throughput improvement and $2.17\times$ inference energy reduction during YOLOv3 and RESNET50 workload on average. In contrast, conventional data orchestration would require more elaborate hardware and control signals to implement im2col in hardware because of the data skew. We have synthesized and conducted place and route for 16×16 systolic arrays based on the novel and conventional orchestrations using ASAP 7nm PDK and found that our proposed approach results in 0.211\% area and 1.6\% power overheads. 
\vspace{-5pt}
\end{abstract}







\maketitle
\vspace{-4pt}
\section{Introduction}
General Matrix multiplication (GeMM) is at the core of current AI applications and in most cases the throughput is limited by how fast it can be executed. General-purpose computers are quite inefficient when it comes to matrix multiplication because of their complex architectures and there have been major research and development efforts on domain-specific architectures to overcome this bottleneck. Hardware accelerators like Graphical Processing Units (GPU) are one of them built according to SIMD (Single Instruction Multiple Data Stream) architecture. GPUs are very popular for their highly parallelization capability which helps to run AI applications faster. Unfortunately, they are extremely power-hungry because of their general-purpose design, and the fact that all communication is done through memory hierarchy\cite{jouppi2017datacenter, 9218732}. 


Systolic array-based architecture has become popular for GeMM because of its simple design, lower memory bandwidth requirement and energy efficiency \cite{xu2023survey,kung1982systolic, 9773232, 9099977, he2020sparse, das2020systolic, cong2018polysa, 9034111} and almost all commerical DNN accelerators use systolic array \cite{9499913, jouppi2017datacenter, choquette2021nvidia, 9116469, nvidia2017nvidia, chen2016eyeriss, chen2019eyeriss, cho2021risa, 8772100, wang2022sadd,venkataramani2021rapid, 9138929, xu2021heterogeneous, 9218732}. Open-source frameworks like Gemmini\cite{genc2021gemmini} also generate systolic array-based designs. Unlike GeMM, Convolution operation needs im2col algorithm to map a 3-D tensor to an equivalent 2-D matrix which enables convolution workload to run in GeMM accelerator. In conventional SA, software-based im2col is used to generate the corresponding matrices which are then stored in SRAM buffers to feed the array during execution. Although it is computationally efficient, it requires significant on-chip memory due to repeated elements in convolution windows. However some of the systolic array based designs are optimized for the convolution workload \cite{xu2021heterogeneous, das2020systolic, lu2017flexflow, liu2020systolic, 9473985}.

 Eyeriss \cite{chen2016eyeriss} addresses this memory disadvantage by proposing a novel dataflow titled row stationary. However, the processing elements  (PE) in Eyeriss must have large Multiply-Accumulate (MAC) units with large register files (RF) compared to those in simpler systolic arrays. This is because RF requires storing a filter row into PE's buffer. Also, IFMAP needs to copy in each set of PEs that are working on producing results for a filter, unlike SA where the same IFMAP copy is reused along all the filters. Like Eyeriss's row stationary dataflow, FlexFlow\cite{lu2017flexflow} has been proposed to address the above mentioned challenge. Both FlexFlow and row stationary dataflows do not follow the fundamental systolic array architecture. Although these designs are efficient in computation, data transmission is dependent on bus communication which limits the frequency of the overall system \cite{xu2021heterogeneous}. Gemmini\cite{genc2021gemmini} introduces optional im2col units which prevent CPU from managing the im2col but this design still suffers from frequent memory accesses. SPOTS\cite{soltaniyeh2021spots} reduces the L2 or OFF-chip memory access but it introduces a large intermediate buffer.  Recently, some work has been done for on-the-fly im2col that replaces software im2col requirement and thus helps to reduce data transfer \cite{fornt2023energy, liu2019usca}. However,  this approach comes with a major hardware overhead including counters, intermediate registers, FIFO, and feed registers, and requires complex control signals.

In this work, we introduce a new systolic array architecture called Axon where we propose a novel in-array data orchestration technique along with a new data transfer from the scratchpad to the array without changing the systolic array's simple design and its reuse capability. In Axon data are fed to PE on the principal diagonal followed by bi-directional propagation, unlike systolic array's uni-directional propagation.
The proposed design provides a direct improvement on runtime for all workloads regardless of their optimal dataflows including output stationary (OS), weight stationary (WS), and input stationary (IS). We present a runtime equation that quantifies the improvements that one can achieve under various scenarios. We also exploit data reuse opportunity when performing convolution by replacing software im2col with a simple 2-to-1 multiplexer (MUX) which requires a significantly smaller 
hardware overhead compared to existing hardware support for im2col \cite{fornt2023energy, liu2019usca}. The proposed redesign of in-array data propagation makes it possible to adopt this data reuse opportunity using a simple 2-to-1 MUX. We also adopted the idea of Zero gating to exploit sparsity in both the input feature map (IFMAP) and FILTERS \cite{fornt2023energy}. Finally, we experimented and validated the runtime improvement through the proposed orchestration, im2col support, and sparsity handling on existing state-of-the-art (SOTA) workloads like transformer (GeMM), conformer (Conv and GeMM) and CNN (conv). Also, we present hardware overhead comparison and power consumption reduction with prior work by performing the full synthesis and physical design process for an application specific integrated circuit (ASIC) implementation on the proposed architecture. 
The key contributions of this paper are as follow:
\vspace{-3pt}
\begin{itemize}
    \item A novel in-array data orchestration in systolic arrays that speeds up GeMM irrespective of dataflow (OS, IS, WS) and workload shapes. Furthermore, Axon achieves 2x speedup on relatively memory bound operations like GEMV and DW-conv due to lower feeding latency and no data skew. 
    \item An im2col hardware support that is simple to program and adds insignificant hardware overhead (using 2-to-1 MUXes only) for exploiting input feature map reuse opportunity which significantly lowers the off-chip memory traffic during convolution
    \item Unified PE design compatible with Axon architecture which is programmable to exploit all the three dataflows (WS/IS/OS)
    \item The capability of exploiting sparsity in IFMAP and FILTERS to reduce power consumption using the Zero gating technique 
    \item Physical design and layout implementation of Axon using a 7-nm FiNFET PDK  for scalability and backend analysis
\end{itemize}
\vspace{-10pt}
\section{Background and Motivation}
\subsection{Conventional systolic array and dataflows}
\begin{figure}[htb]
    \centering
    \vspace{-10pt}
    \includegraphics[width=\linewidth]{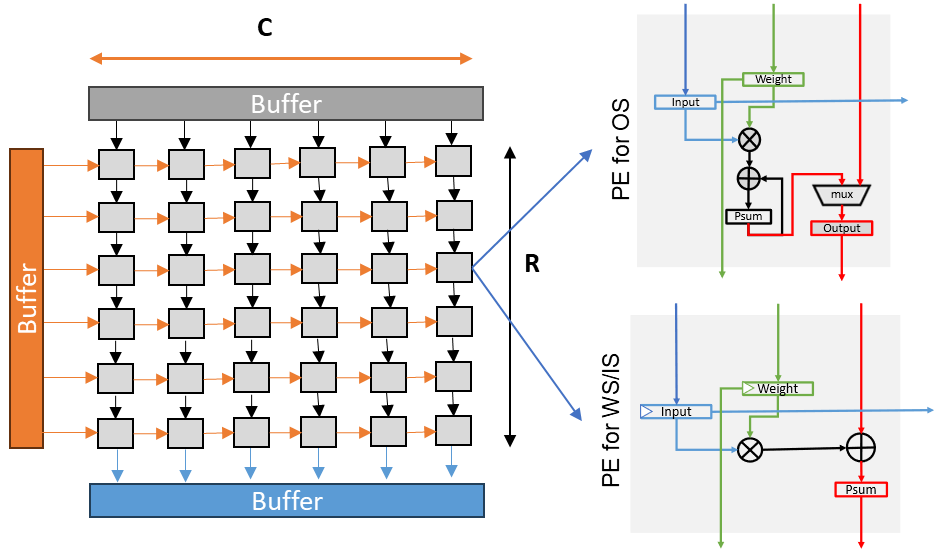}
    \vspace{-25pt}
    \caption{Data feeder from the buffer and in-array dataflow in conventional systolic array with PE's architecture}
    \label{fig:fig2}
    \vspace{-10pt}
\end{figure}

Dataflows in a systolic array define how the elements of operand matrices and partial sums propagate inside the array. 
In the OS dataflow, partial sums remain stationary while operands from both input and weight propagate through the array in a systolic order. At the end of the operation, outputs are extracted from the array. In WS, weights are pre-filled inside the array and remain stationary until the input propagates through the whole systolic array. In each cycle, partial sums are generated in each PE and propagate downward to be added in the next cycle with another PE-generated partial sum. Finally outputs are collected from the bottom row of the array. IS is similar to WS with the only difference being that inputs instead of weights are pre-filled in the array and remain stationary. Fig.~\ref{fig:fig2} demonstrates in-array dataflows in the conventional systolic array, in which the input feature map/one operand matrix is loaded from the input feature map buffer to the systolic array through the leftmost PEs and propagates from left to right. Weights/another operand matrix are loaded from the weight buffer through PEs in the top rows and propagate from top to bottom. Each PE in the array is responsible for one MAC operation and propagates inputs and outputs based on the dataflow type. 
\vspace{-10pt}
\subsection{Runtime Modeling}
\begin{figure}[htb]
    \centering
    \vspace{-10pt}
    \includegraphics[width=\linewidth]{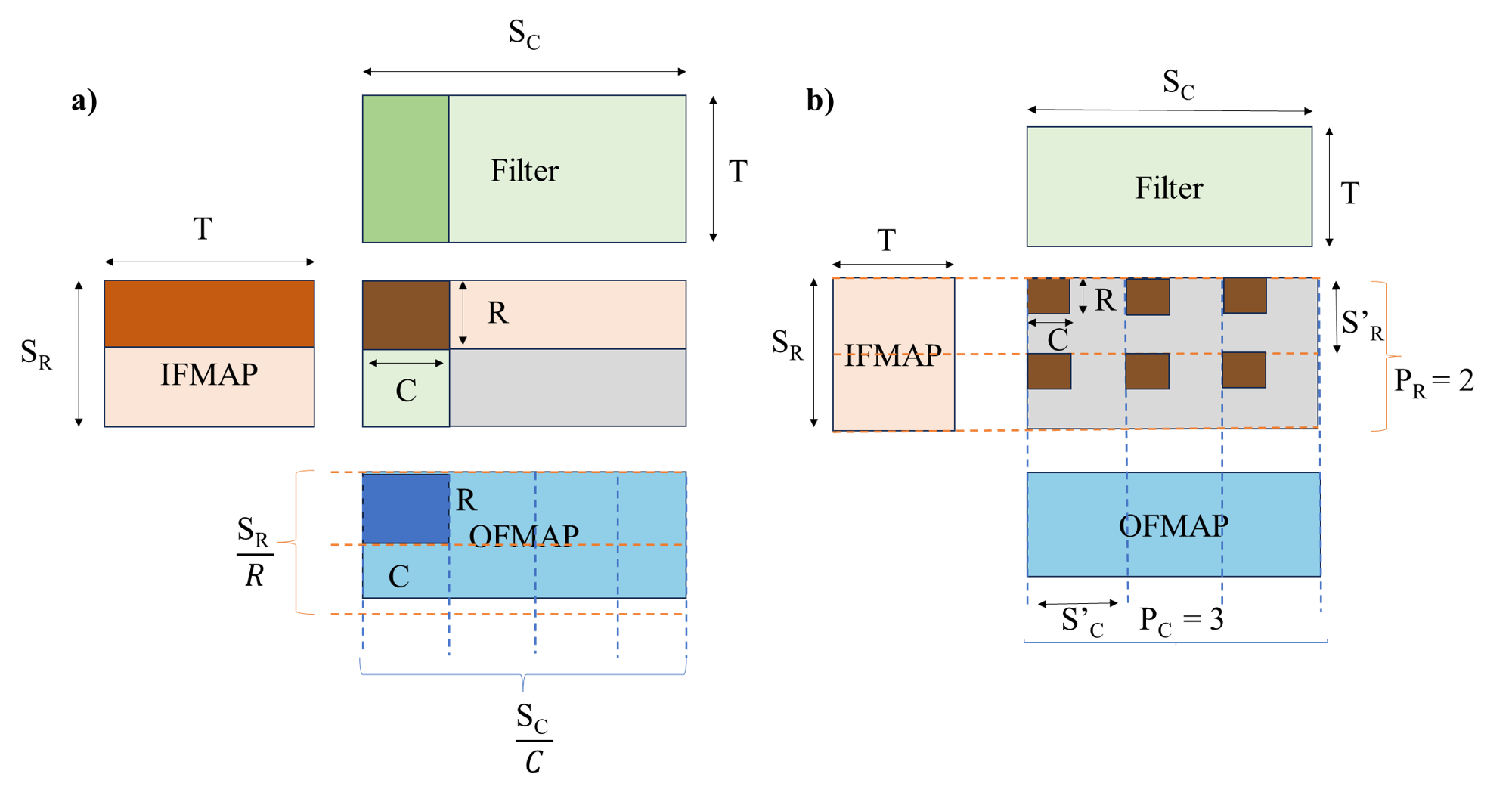}
    \vspace{-20pt}
    \caption{a) Scale up (left) b) Scale out (right). In scale up one large monolithic array is used whereas in scale out multiple systolic arrays are used to generate output.}
    \label{fig:fig6}
    \vspace{-15pt}
\end{figure}

SCALE-SIM\cite{samajdar2020systematic} developed an analytical model for systolic array runtime,  which is applicable for all the three dataflows discussed above. According to the model, for the operand matrices of dimensions $S_R \times T$ and $T\times S_C$ respectively the runtime is as follows:
\begin{equation}
    \begin{aligned}
        \tau=2 S_R+S_c+T-2
    \end{aligned}
    \label{eq1}
\end{equation}
where $S_R$ and $S_C$ are the spatial dimensions along which computation is mapped, and $T$ represents the corresponding temporal dimension. The original operand matrices are projected into the available spatio-temporal dimensions. For example, during the multiplication of two matrices of shapes $M \times K$ and $K \times N$, the dimension $M$ is mapped to $S_R$, dimension $N$ is mapped to $S_C$ and the dimension $K$ to $T$ for OS dataflow. Mapping for three dataflows are summarized in Table.\ref{tab:dfmapping}. 
\begin{table}[h]
\centering
\vspace{-5pt}
\begin{tabular}{ll}
\toprule
Dataflow & Mapping \\
\midrule
OS  & ($S_R = M, S_C = N, T = K$) \\
WS  & ($S_R = K, S_C = M, T = N$) \\
IS  & ($S_R = K, S_C = N, T = M$) \\
\bottomrule
\end{tabular}
\caption{Mapping of GEMM dimension along the array during different dataflows}
\vspace{-30pt}
\label{tab:dfmapping}
\end{table}
However, Large GEMM problems are managed on smaller systolic arrays through tiling in scale-up (one large array) or scale-out (multiple smaller arrays) as depicted in Fig.~\ref{fig:fig6}. Runtime equation for the scale up and scale out is modified as Eq.~\ref{RuntimeScaleup} and \ref{RuntimeScaleout}, respectively, 

\begin{equation}
    \begin{aligned}
        \tau_{scaleup} = (2R + C + T -2) * (S_R / R) * (S_C / C)
    \end{aligned}
    \label{RuntimeScaleup}
\end{equation}
\begin{equation}
    \begin{aligned}
        \tau_{scaleout} = (2R + C + T -2) * (S'_R / R) * (S'_C / C)
    \end{aligned}
    \label{RuntimeScaleout}
\end{equation}
where $R$ and $C$ are the number of rows and columns of the systolic array, $S'_R$ and $S’_C$ are $(S_R/P_R)$ and $(S_C/P_C)$, respectively,  $P_R$ is the partition number across the rows, and $P_C$ is the partition number across the columns. If we break down the runtime equation, we find the following three components: 
\begin{itemize}
    \item Time for both operands to reach the farthest PE with respect to the feeder PEs, $R + C -2$
    \item Number of multiplications each PE performs, $T$
    \item Readout from the array, $R$ 
\end{itemize}

For any dataflow (OS, WS, and IS) the number of multiplications each PE performs is determined by the temporal dimension, $T$. The readout time is determined by the number of rows of the systolic array, $R$.  However, the first component is determined by the in-array data propagation direction. For the conventional systolic array, it is the Manhattan distance where the distance is defined by the distance from the vertical and horizontal axes. For the farthest PE (bottom right corner) in the array, the distance is $(R + C - 2)$. 

\begin{figure}[htb]
    \centering
        
    \includegraphics[width=\linewidth]{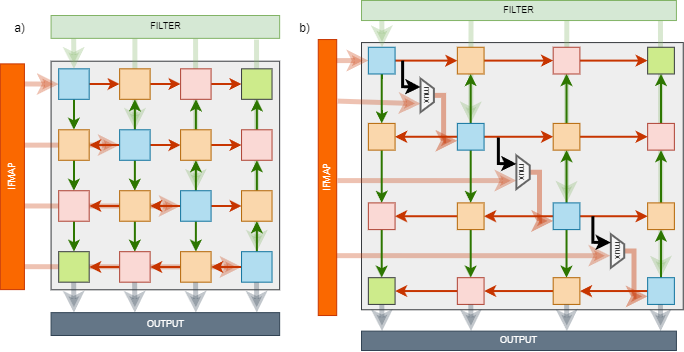}
        \vspace{-15pt}
    \caption{a) Axon's in-array data orchestration. Thick semitransparent arrows indicate data movement into the systolic array from buffers and thin solid arrows indicate data movement inside the array among the PEs. The same colors on PEs represent operands' arrival at the same cycle whereas PEs on the principal diagonal receive the operands on the first cycle directly from the buffers. b) im2col implementation. Note that, each MUX allows feeder PEs to receive data either from buffer or from immediate PE on the diagonal.}
    \vspace{-5pt}
    \label{fig:fig3}
\end{figure}
\section{Axon Data Orchestration Strategy}
\begin{figure}[tb]
    \centering
    \vspace{-5pt}
    \includegraphics[width=\linewidth]{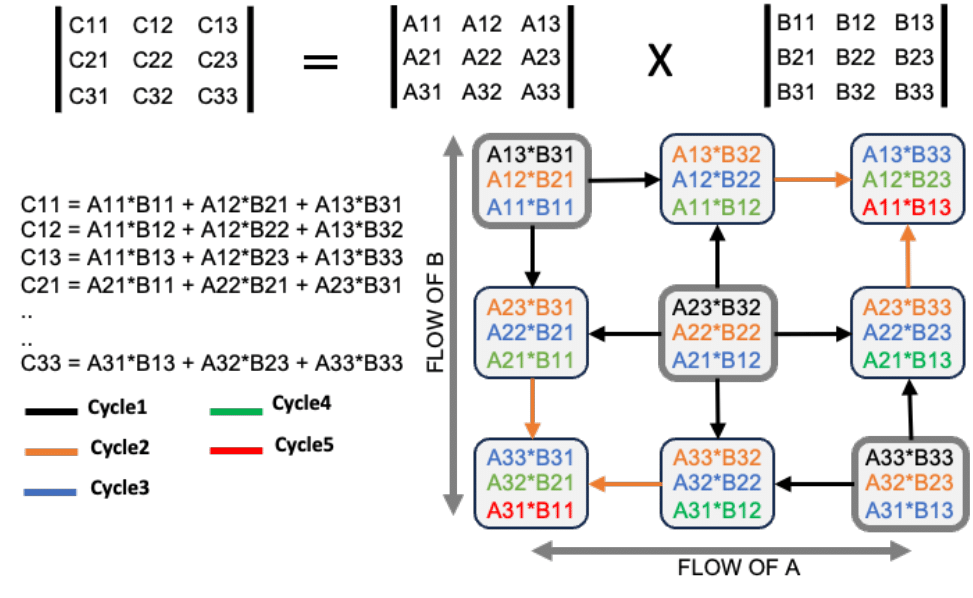}
    \vspace{-20pt}
    \caption{Simple $3\times3$ GeMM example that validates Axon data orchestration. Partial products labeled with the same colors indicate that they are generated in the same cycle. All operands are fed to SA through PE on the principal diagonal.} 
    \label{fig:fig4}
    \vspace{-15pt}
\end{figure}

\begin{figure}[htb]
    \centering
    \includegraphics[width = \linewidth]{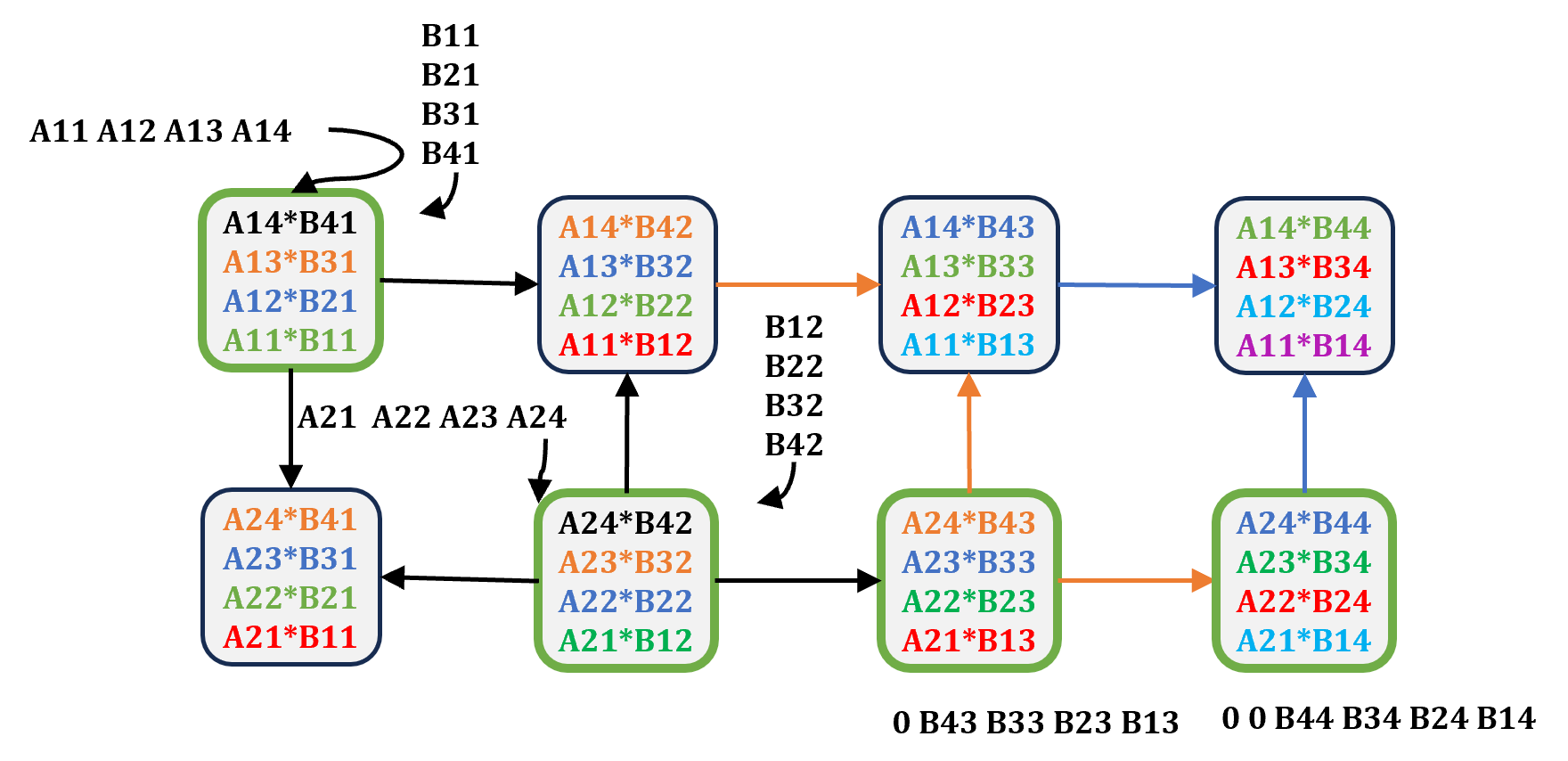}
    \vspace{-25pt}
    \caption{Axon data orchestration in the rectangular systolic array. The columns that do not have any PEs on the principal diagonal will be fed through PEs at the bottom of the array with zero padding based on the distance. Notice third column is fed with zero-padded by one and the fourth column is fed with zero-padded by two. }
    \vspace{-10pt}
    \label{fig:rectAxon}
\end{figure}

In the conventional orchestration, it takes  $(R + C - 2)$ cycles for the operands to reach the farthest PE. The Axon data orchestration shown in Fig.~\ref{fig:fig3} improves this portion of the runtime by feeding the operand matrices to the systolic array through the PEs on the principal diagonal which we call feeder PEs (semi-transparent thick arrows). Once data is inserted into the systolic array, data movement inside the array is orchestrated according to Fig.~\ref{fig:fig3} (solid-thin arrows) to perform the matrix multiplication. The feeder PEs except the two corner PEs transfer the operands in two directions, unlike other PEs on the array. Filter elements are allowed to propagate towards both adjacent upward and downward PEs whereas IFMAP elements are allowed to propagate towards both adjacent right and left PEs. Other PEs in the array propagate the operand in the same direction as they receive data. Note that in this way, we do not need any extra hardware. We only need to rearrange the in-array dataflow direction and feed the array from the PEs on the principal diagonal. Moreover, Unlike conventional SA Axon does not need to stream the operand matrices in a skewed manner which increases PE utilization. Fig.~\ref{fig:fig4} shows a toy example to demonstrate the Axon data orchestration through a two $3 \times 3$ matrix multiplication. 
While systolic arrays are often square, the Axon data orchestration can also be extended for rectangular systolic arrays and reduce the runtime as shown in Fig.~\ref{fig:rectAxon}. For the columns that do not have any PEs on the principal diagonal, operands are fed from the bottom PE of the column with spatially skewed data similar to a conventional systolic array which ensures accurate data orchestration for the PEs. 
\begin{table}[h]
\centering
\begin{tabular}{lll}
\toprule
Dataflow & Systolic array & Axon \\
\midrule
OS & $2M + K + N - 2$ & $\max(M, N) + M + K - 1$ \\
WS & $2K + M + N - 2$ & $\max(M, K) + K + N - 1$ \\
IS & $2K + M + N - 2$ & $\max(N, K) + K + M - 1$ \\

\bottomrule
\end{tabular}
\caption{Runtime for SA and Axon}
\vspace{-15pt}
\label{tab:runtimeSummary}
\vspace{-20pt}
\end{table}
\subsection{Runtime modeling}
\begin{figure}[htb]
    \centering
    \includegraphics[width=\linewidth]{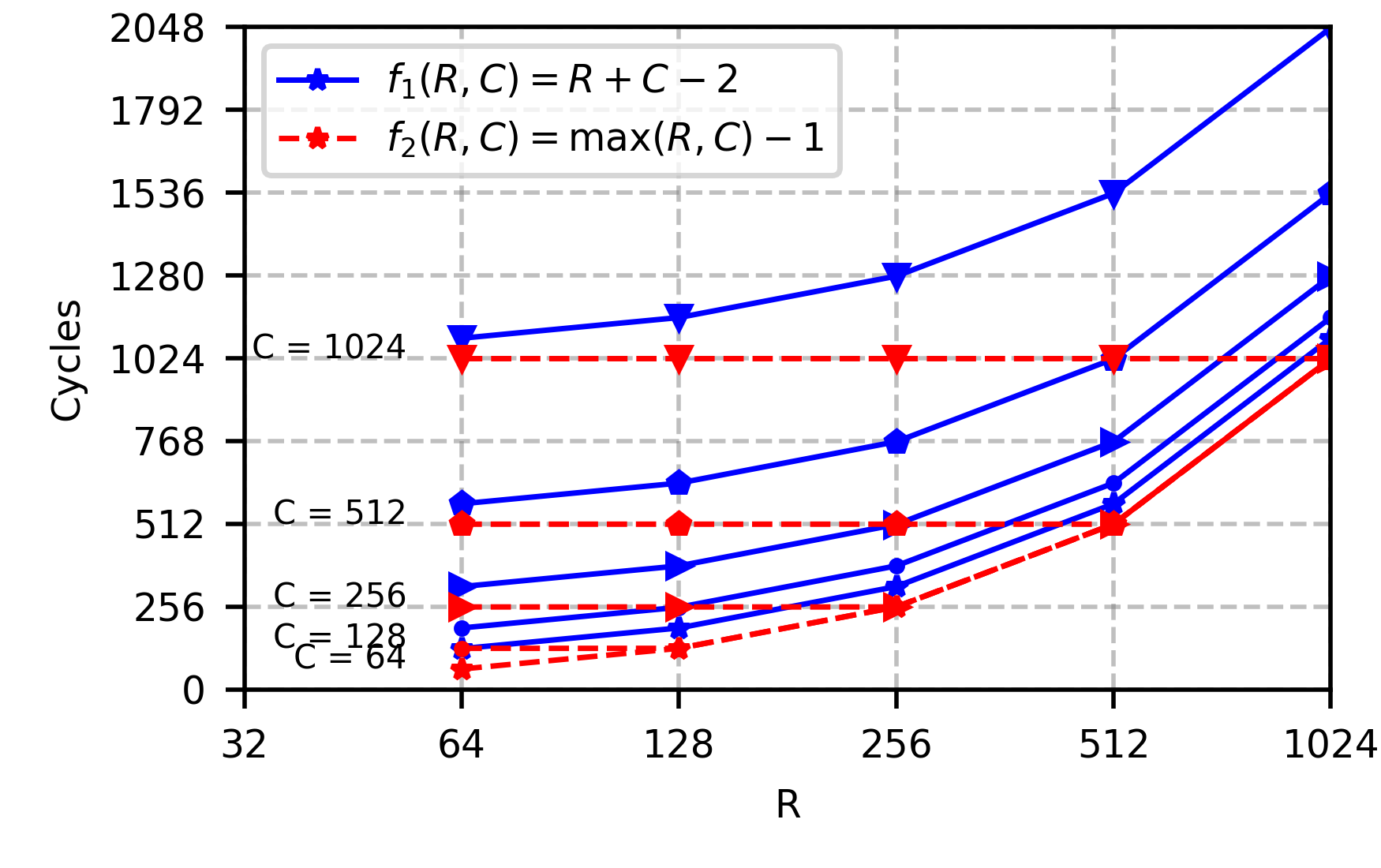}
    \vspace{-30pt}
    \caption{Runtime factor (cycles) responsible for feeding operand to the farthest PE. $f_1(.) $ represents conventional systolic array, $f_2(.)$ represents Axon data orchestration.}
    \label{fig:fig5}
    \vspace{-10pt}
\end{figure}

The $2^{nd}$ and $3^{rd}$ terms in the runtime equation presented by Samajdar et al \cite{samajdar2020systematic} remain unchanged in this orchestration while the $1^{st}$ term is changed to $max(R, C)-1$. For a square systolic array$(R=C)$, it is simply $(R-1)$ which is half of the $(2R -2)$ for the conventional systolic array. The improvement for non-square arrays is smaller but is always greater than 1. Table.~\ref{tab:runtimeSummary} summarizes the runtime for various dataflows w.r.t their operand matrix shape considering $S_R = R$ and $S_C = C$. Fig.~\ref{fig:fig5} shows that for any shape of the array $(R, C)$ the factors that determine the time it takes for the operands to reach the farthest PE are always lower in Axon data orchestration. For example, in a systolic array of shape $(256, 256)$  the compute time required to reach the farthest PE is reduced from 510 cycles to 255 cycles. However, the overall runtime improvement for a scale-out design with smaller arrays is limited by two other factors (i.e. Temporal dimension length and Readout time) according to Amdahl's law.
\vspace{-5pt}
\subsection{Axon's hardware support for im2col}
\begin{figure}[htb]
    \centering
    \includegraphics[width=\linewidth]{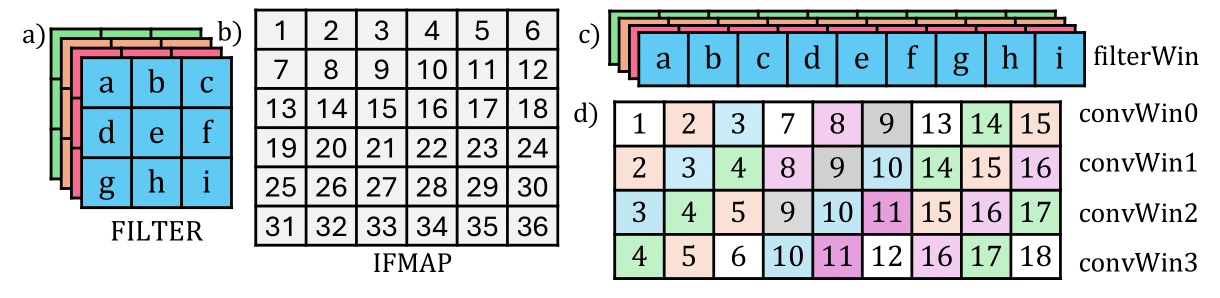}
    \caption{Im2col on a) FILTER of shape (3x3) and b) IFMAP of shape (6x6). After im2col each filter is flattened like c) and each conv window is flattened as shown in d). (Only 4 conv windows have been shown related to producing $1^{st}$ row of the output).}
    \label{fig:convReuseOp}
    \vspace{-15pt}
\end{figure}

 During im2col, IFMAP is converted to conv windows depending on the shape of the FILTER. Each conv window is responsible for generating one element of OFMAP. In the example shown in Fig.~\ref{fig:convReuseOp}, Filter and IFMAP are of shape $3\times3$ and $6\times6$ respectively. Thus OFMAP shape is $4\times4$. In other words, there are in total 16 conv windows corresponding to 16 elements of the $4\times4$  OFMAP. In the example only 4 conv windows correspond to $1^{st}$ row of OFMAP has been shown. Note that in conv windows, there are 18 unique elements and the remaining 18 elements have been repeated (50\% repetition) in a pattern that can be formulated in terms of the FILTER length, $n$. The number of common elements in consecutive conv windows is $n(n-1)$, which is $3\times2$ or $6$ for the demonstrated example. Also between each adjacent conv window, there is a cyclic pattern with a period of $n$ where there are $n-1$ common elements. This repetition becomes more prominent for large FILTER and IFMAP shapes which results in excessive memory traffic and a need for either a large on-chip memory or expensive DRAM access. In Axon, we leverage this repetition pattern to reuse the elements directly from the PEs instead of DRAM access or having a large on-chip memory. 

Fig.~\ref{fig:fig3}(b) demonstrates the proposed Axon's simple im2col hardware support where each PE on the principal diagonal can be fed with IFMAP elements either directly from IFMAP SRAM buffers or from the adjacent top PE on the principal diagonal using multiplexers (MUX), where the control signal is 0 for 1 cycle and 1 for the other $(n-1) $ cycles. With this configuration, during the $1^{st}$ cycle, all the 4 elements (the rightmost element from each row of IFMAP conv window matrix in Fig.~\ref{fig:convReuseOp}(d)) will be loaded from the SRAM buffer through feeder PEs of the array. In the $2^{nd}$ cycle, only the row with convWin0 will load the next element from SRAM buffer while all the other three conv windows will get their $2^{nd}$ elements from the adjacent top feeder PEs via mux i.e. the feeder PE of convWin1 will get element from the feeder PE of convWin0. Similarily the feeder PE of convWin2 will get the element from convWin1's feeder PE and so on. In the $3^{rd}$ cycle, convWin1, convWin2, convWin3 will get their element from the top adjacent feeder PE just like $2^{nd}$ cycle. The same dataflow will repeat from the $4^{th}$ cycle since the FILTER length is 3). Hence, Axon can readily exploit the reuse and just load the PEs with elements from immediate feeder PE instead of loading from memory for $(n-1)$ cycles out of $n$ periodic cycles. Thus, we can avoid the high memory traffic imposed by software-based im2col with a minimal hardware overhead unlike hardware support proposed in prior works. The proposed novel data orchestration makes this simple scheme possible. In our approach data is loaded in an ordered fashion to the feeder PEs from memory in contrast to the conventional orchestration where the data is skewed.

\begin{figure}[htb]
    \centering
    \vspace{-10pt}
    \includegraphics[width=\linewidth]{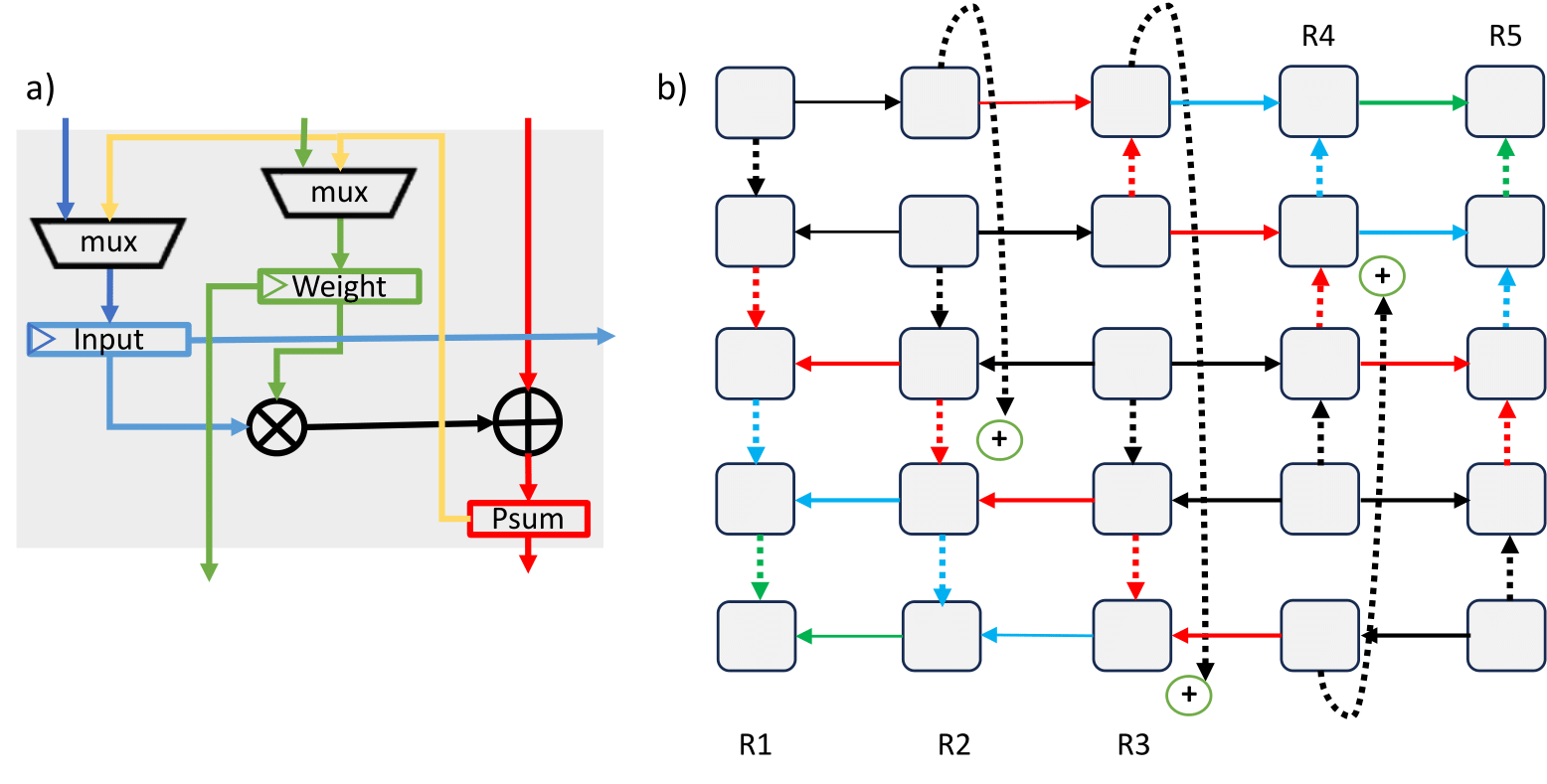}
    \vspace{-25pt}
    \caption{a) IS/WS PE for Axon, b) Partial sums are coordinated using bypassing at the same stage apart from the feeder PE in the column to avoid partial sum data corruption with different output elements. Identical arrow color represents same cycles during the dataflow. }
    \vspace{-15pt}
    \label{fig:fig11}
\end{figure}

\vspace{-10pt}
\section{PE design for Axon}
PEs are commonly designed based on the dataflow\cite{xu2023survey}, and the design of PEs is slightly different for IS and WS than the OS. Fig.~\ref{fig:fig2} depicts the standard design for PEs of OS and IS/WS. For Axon data orchestration, the only change in the design of the PEs is in terms of data propagation. We will discuss the design of PEs for each dataflow especially for WS/IS due to their design challenges for Axon architecture in the following section.  
\vspace{-10pt}
\subsection{Output Stationary dataflow}
PE design of Axon OS dataflow is identical to the conventional design as depicted in Fig.\ref{fig:fig2}. In Axon, PE on principle diagonal except the ones at the two corners propagate data in both directions, unlike the PE in conventional design. So the only change in hardware is the addition of an interconnect to allow bi-directional propagation of data. The PE also integrates the zero-gating approach presented in\cite{shao2023efficient} where MAC operation is simply skipped if zero exists in either IFMAP or FILTER operand. This helps to reduce power during sparse GEMM.

\vspace{-10pt}
\subsection{Input/Weight Stationary dataflow}
Axon data orchestration is straightforward to implement for OS dataflow. IS and WS dataflows; however, raise the two following challenges:


\vspace{-5pt}
\subsubsection{Preloading}
From Fig.\ref{fig:fig3} we observe that data is fed from the buffer to the systolic array through the diagonal elements and the flow is in both directions (top-bottom, left-right). Unlike OS, we require preloading weight/input for WS/IS dataflow. The loading takes $S_R$ cycles. Loading from the buffer to the array according to the proposed architecture becomes faulty because of dataflow in both directions. To solve the problem, we have utilized an output interconnect which is used to propagate the output vertically (Fig.\ref{fig:fig11}(a) yellow route). Two additional 2-to-1 MUXes are required to determine the target buffer based on the dataflow. Note that the in-array dataflow direction during computation remains unchanged like the OS dataflow.  
\begin{figure}[htb]
    \centering
    \vspace{-10pt}
    \includegraphics[width=0.8\linewidth]{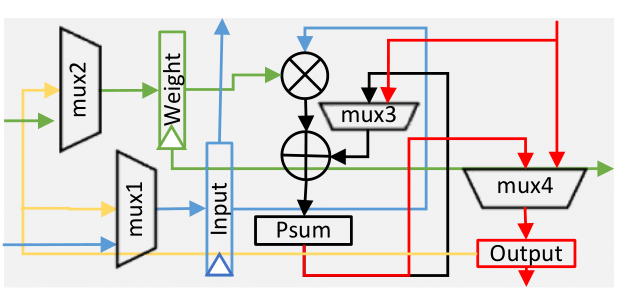}
    \vspace{-15pt}
    \caption{Axon Unified PE for OS, IS, and WS.}
    \vspace{-15pt}
    \label{fig:fig12}
\end{figure}
\vspace{-5pt}
\subsubsection{Partial sum synchronization}
Another challenge that comes from implementing the Axon data orchestration in WS/IS is partial sum synchronization. In a conventional systolic array, the operand elements propagate from top to bottom and left to right and the partial sum is also calculated and propagated downward. In each stage, a partial sum gets added to the previous stage's total sum along a column and finally leaves the array from the bottom row PEs as the output. Thus, it ensures that the correct partial sums are added in the right order. However, in the proposed architecture the operand elements propagate in both directions; hence, the partial sums corresponding to the same output element are generated in parallel. They must be collected for accurate output generation before leaving the array. To address the issue we use the bypass and add approach which adds two portions of the final output elements separated by the PEs on the principal diagonal on that column. The PEs on the principal diagonal propagate operands in both directions but propagate their output in only one direction (either to the bottom or to the top). Fig.\ref{fig:fig11}(b) depicts the bypassing for IS/WS dataflow where the dashed arrows represent the direction of partial sums propagation and the solid arrows indicate the direction of the operand propagation. Remember that in IS/WS one operand (either IFMAP or FILTER operands) is preloaded and kept stationary where the other operand is allowed to propagate. Bypassing allows to avoid output being corrupted and ensures output is collected without pushing stalls.

\vspace{-5pt}
\subsection{Axon Unified PE design for OS, IS/WS dataflow}
In this section, we introduce a unified PE design as depicted in Fig.~\ref{fig:fig12} that is programmable to switch the data to any of the three dataflows for the Axon data orchestration. Two Multiplexers (MUX1 and MUX2) are used to select the target buffer based on WS/IS dataflow during preloading. They forward data from the buffer to the input/weight buffer through the output path (yellow route). For the WS/IS dataflows, MUX3 will forward the previous partial sum to the adder and then forward the final sum to the output buffer to propagate to another PE's MUX to be summed in the next cycle. For the OS, the black route will be active. Through MUX3 and Psum, the previous partial sum buffered in Psum will be summed with the newly generated partial sum, and finally, through MUX4 the result is written to the output buffer.

\begin{table}[h]
\centering
\scriptsize
\setlength{\tabcolsep}{2pt}
\renewcommand{\arraystretch}{0.9}
\begin{tabular}{lrrr|lrrr}
\toprule
Workload & M & K & N & Workload & M & K & N \\
\midrule
TFO & 31999 & 84 & 1024 & NCF1 & 256 & 2048 & 256 \\
TF1 & 84 & 4096 & 1024 & DB0 & 1024 & 50000 & 16 \\
GNMTO & 128 & 4096 & 2048 & DB1 & 35 & 2560 & 4096 \\
GNMT1 & 2048 & 32 & 4096 & Resnet50\_0\_conv2d & 64 & 147 & 62500 \\
GPT3\_0 (matmul0) & 1024 & 1024 & 80 & Resnet50\_1\_conv2d & 512 & 4608 & 676 \\
GPT3\_1 (matmul1) & 1024 & 2560 & 7680 & YOLO\_v3\_0\_conv2d & 64 & 288 & 42436 \\
GPT3\_2 (addmm) & 1024 & 2560 & 10240 & YOLO\_v3\_1\_conv2d & 128 & 576 & 10404 \\
GPT3\_3 (lmhead) & 1024 & 2560 & 50257 & GEMM\_0 & 128 & 10 & 128 \\
NCF0 & 2048 & 128 & 1 & GEMM\_1 & 2048 & 10 & 2048 \\
& & & & GEMM\_2 & 1024 & 1024 & 128 \\
& & & & GEMM\_3 & 64 & 2560 & 2560 \\
\bottomrule
\end{tabular}
\caption{Values of M, K, and N for different workloads}
\label{tab:data}
\vspace{-20pt}
\end{table}

\vspace{-10pt}
\section{Evaluation}
In this section, we evaluate Axon architecture against conventional systolic array and configurable multi-directional systolic array(CMSA) \cite{xu2021configurable}. CMSA adds an additional datapath to the systolic array to improve computing efficiency. 
We used the analytical model adopted from SCALEsim\cite{samajdar2020systematic} for SA runtime calculation and the analytical model from the CMSA paper. For im2col hardware support we compared with Sauria\cite{fornt2023energy}. Sauria has a data feeder to support on-the-fly im2col. TSMC 45nm PDK and ASAP 7nm PDK\cite{clark2016asap7} have been used for the RTL synthesis and PnR. We used Synopsys VCS and DCSHELL for RTL design and functional verification. We evaluated the speedup of Axon architecture over the baselines for GEMV, DW-Convolution, GEMM, and Convolution workloads. Workloads from Transformers (e.g. GPT3)\cite{brown2020language, ashish2017attention}, CNN(e.g. RESNET, Efficientnet, Mobilenet and YOLO V3), Conformer and Sparse GEMM have been used\cite{he2016deep, tan2019efficientnet, howard2017mobilenets, redmon2018yolov3, gulati2020conformer}. Table.~\ref{tab:data} summarizes the M, K, and N values corresponding to GEMM and Conv (mapped to GEMM) workloads. For comparison, we used scale-up to calculate the run time for the experimental demonstration. The run-time improvement in scale-up due to the proposed orchestration will be reflected linearly in the scale-out as well.
\begin{figure}[htb]
    \centering
    \includegraphics[width=\linewidth]{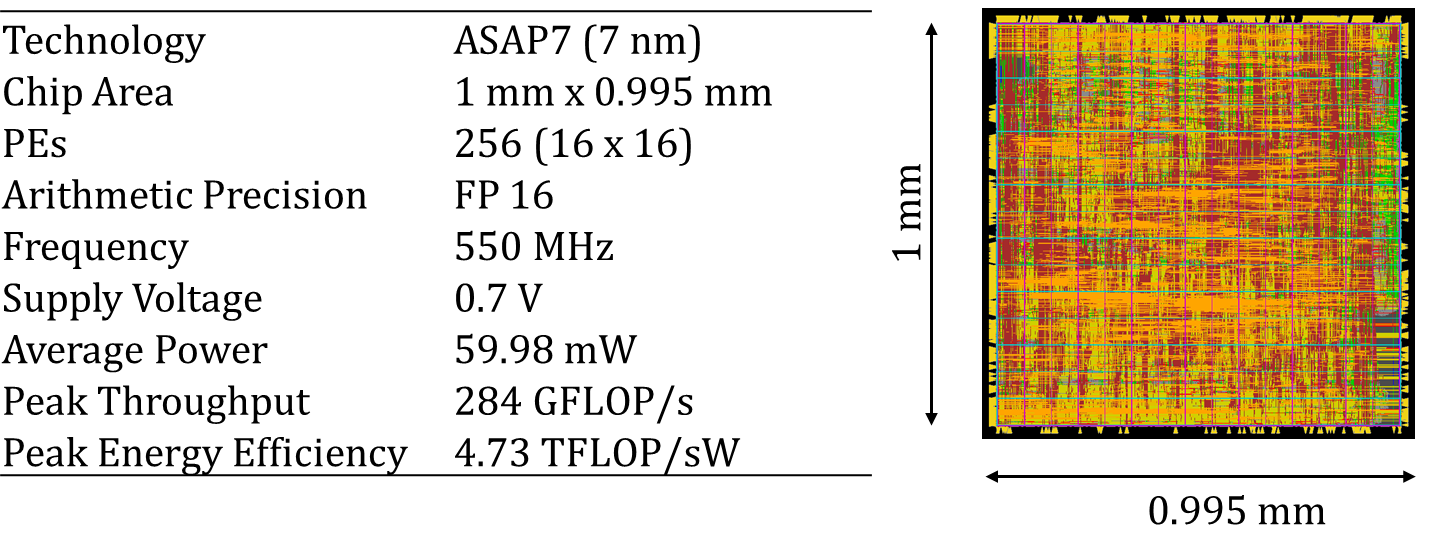}
    \vspace{-25pt}
    \caption{Implemented Axon with im2col support Specifications (left). Floorplan (post PnR) of the chip.}
    \vspace{-12pt}
    \label{fig:Powerreport}
\end{figure}
\vspace{-10pt}
\subsection{Hardware implementation}
We designed and verified Axon architecture with OS dataflow of shape $16\times16$ that includes the proposed im2col hardware support and zero gating module. The OS dataflow is used because it offers high IFMAP and filter reuse opportunities\cite{deng2020model}. We used a simplified version of the FPnew floating-point unit from the open-source parallel ultra-low power (PULP) platform\cite{mach2020fpnew} to implement FP16 MAC unit. Fig.~\ref{fig:Powerreport} summarizes the specifications. We found $0.9992m^2$ of Si area for conventional SA and $0.9931mm^2$ for Axon. The slight reduction in the area is due to buffer sharing between two adjacent PEs of the feeder PE on the principal diagonal. The Input and weight buffers of PEs can be shared between two PEs horizontally and vertically separated by PE on principal diagonal respectively as they receive the same data in the same cycle and are at the same distance from the feeder PE. After adding Im2col hardware support, the Si area becomes increases by only $0.2\%$ ( total area of $0.9951mm^2$). The total power increases by 1.6\% (59.98mW) with im2col support compared to SA (59.88mW) which indicates that power change is insignificant compared to the conventional SA. 
\begin{figure}[hb]
    \centering
    \includegraphics[]{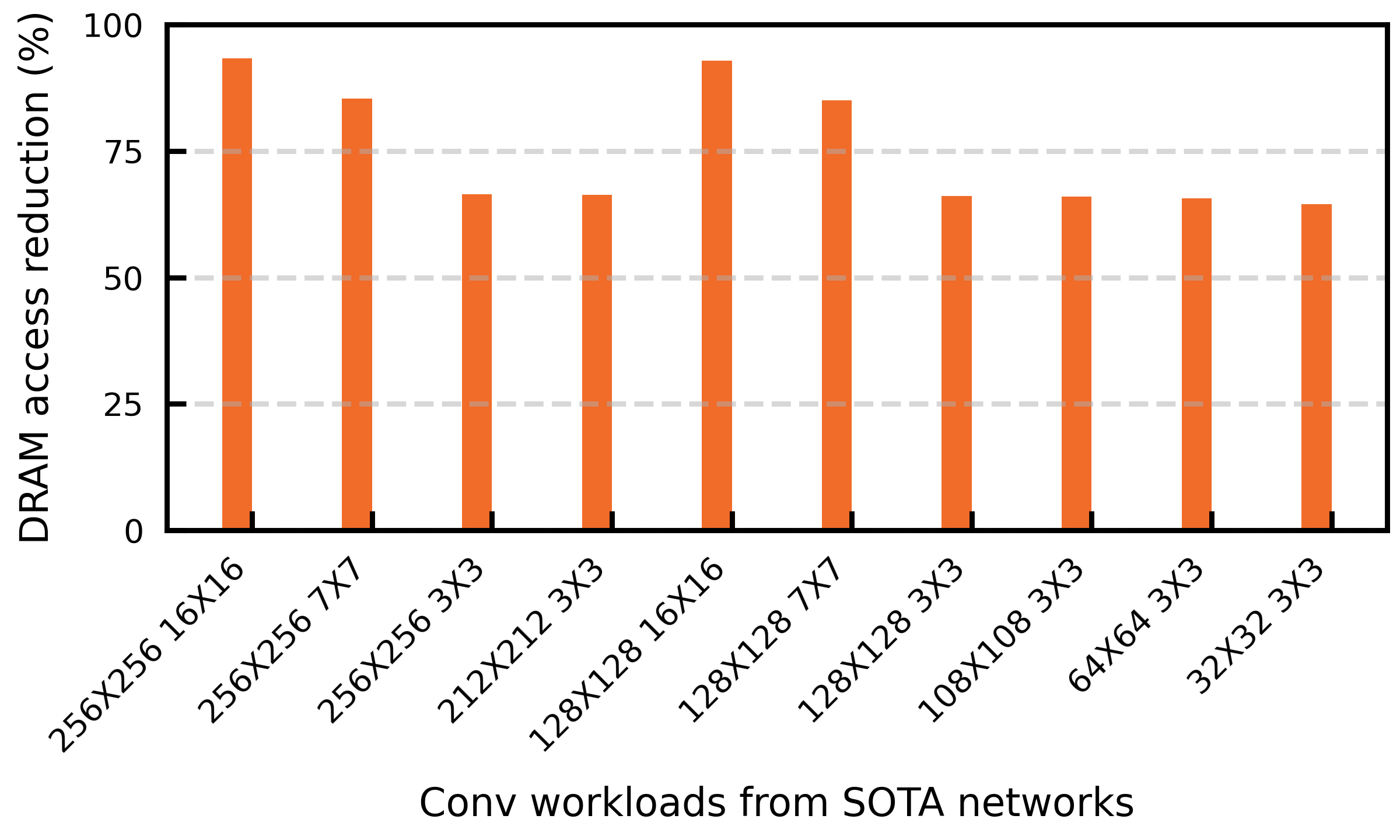}
    \vspace{-10pt}
    \caption{Memory access reduction using proposed on chip HW support for im2col for different IFMAP and kernel shapes adopted from SOTA neural networks }
    \vspace{-15pt}
    \label{fig:memAccessRedPerf}
\end{figure}
\vspace{0pt}
\subsection{Results}
\vspace{-5pt}
\subsubsection{Comparison with SA}
\begin{figure}[tb]
    \centering
    \vspace{-12pt}
    \includegraphics[width=\linewidth]{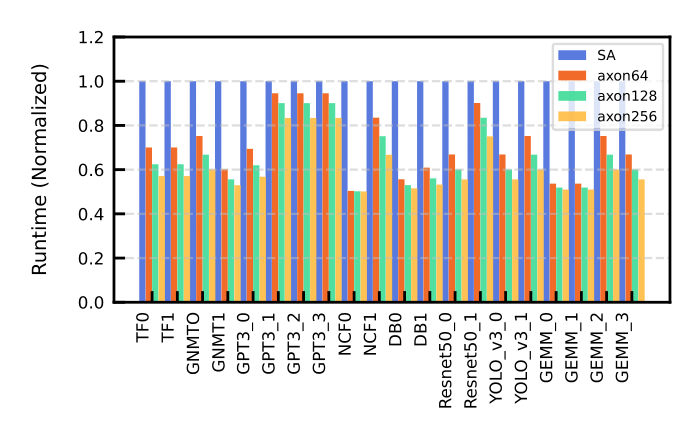}
    \vspace{-30pt}
    \caption{Runtime improvement evaluation on GEMM and Conv workloads}
    \label{fig:GEMMandConvPerf}
    \vspace{-15pt}
\end{figure}

Fig.~\ref{fig:GEMMandConvPerf} depicts the performance comparison of the systolic array with Axon for GEMM and Convolution workloads as summarized in Table.~\ref{tab:data} for various array shapes. We have normalized the runtime (cycles) w.r.t corresponding systolic array runtime. On average, we observe Axon offers $1.47\times$ speedup over the SA for an array shape of $64\times64$. For larger arrays Axon performs even better for most workloads. However, for some workloads (e.g. NCF0 and DB0) for which the runtime is limited by the temporal dimension (e.g. NCF0 and DB0), scaling up doesn't help to improve performance. We achieve an average of $1.76\times$ speedup over conventional systolic array for the array shape of $256\times256$. Fig.~\ref{fig:DW-MV perf(type2)} depicts the performance comparison on depthwise convolution(DW-Conv) and general matrix-vector multiplication (GEMV) workloads between SA and Axon. It shows that Axon is suited for low AI (Arithmetic Intensity) operations like GEMV and DW-Conv because of its bidirectional data propagation. We achieve an average speedup of $1.8\times$. We have used the simple integrated zero gating technique presented in \cite{fornt2023energy} to lower the power consumption by leveraging the sparsity and were able to achieve a 5.3\% total power reduction for the case of 10\% sparsity. Fig.\ref{fig:memAccessRedPerf} demonstrates the capacity of the proposed on-chip im2col to reduce memory access for various convolution workload shapes. We observe that the memory access can be reduced by more than $60\%$  by the on-chip im2col hardware for workloads generally used in SOTA neural networks.  The hardware overhead for the im2col is $<1\%$ found by performing PnR of a $16\times16$ array. We calculate the energy consumption reduction due to lowering DRAM access for Resnet50 and YOLOv3 models. Memory access (for conv layer only) reduced from $261.2MB$ to $153.5MB$ for Resnet50 and from $2540MB$ to $1117MB$ for YOLOv3. Considering LPDDR3 memory where energy cost is 120pJ/byte as found by \cite{chandrasekar_drampower} we find that the inference energy is reduced by $12mJ$ for Resnet50 and $170mJ$ for YOLOv3. We considered 32-bit-wide LPDDR3 DRAM memory at $800 MHz$ with a $6.4 GB/s$ maximum bandwidth and found about $1.25\times$ speedup due to lower memory traffic enabled by im2col support which is comparable as reported by \cite{fornt2023energy}. But the proposed im2col area overhead is only $0.2\%$ where the feeder network of \cite{fornt2023energy} is $4\%$. 
\vspace{-5pt}
\subsubsection{Comparison with CMSA}
\begin{figure}
    \centering
    \includegraphics[width=\linewidth]{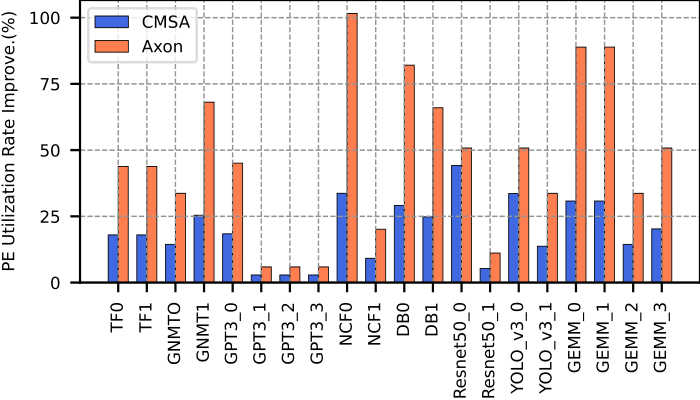}
        \vspace{-20pt}
    \caption{PE utilization rate improvement over conventional systolic array for CMSA and Axon architecutre.}
    \label{fig:UR_imp_comp}
    \vspace{-20pt}
\end{figure}

\begin{figure}[htb]
    \centering    \includegraphics[width=\linewidth]{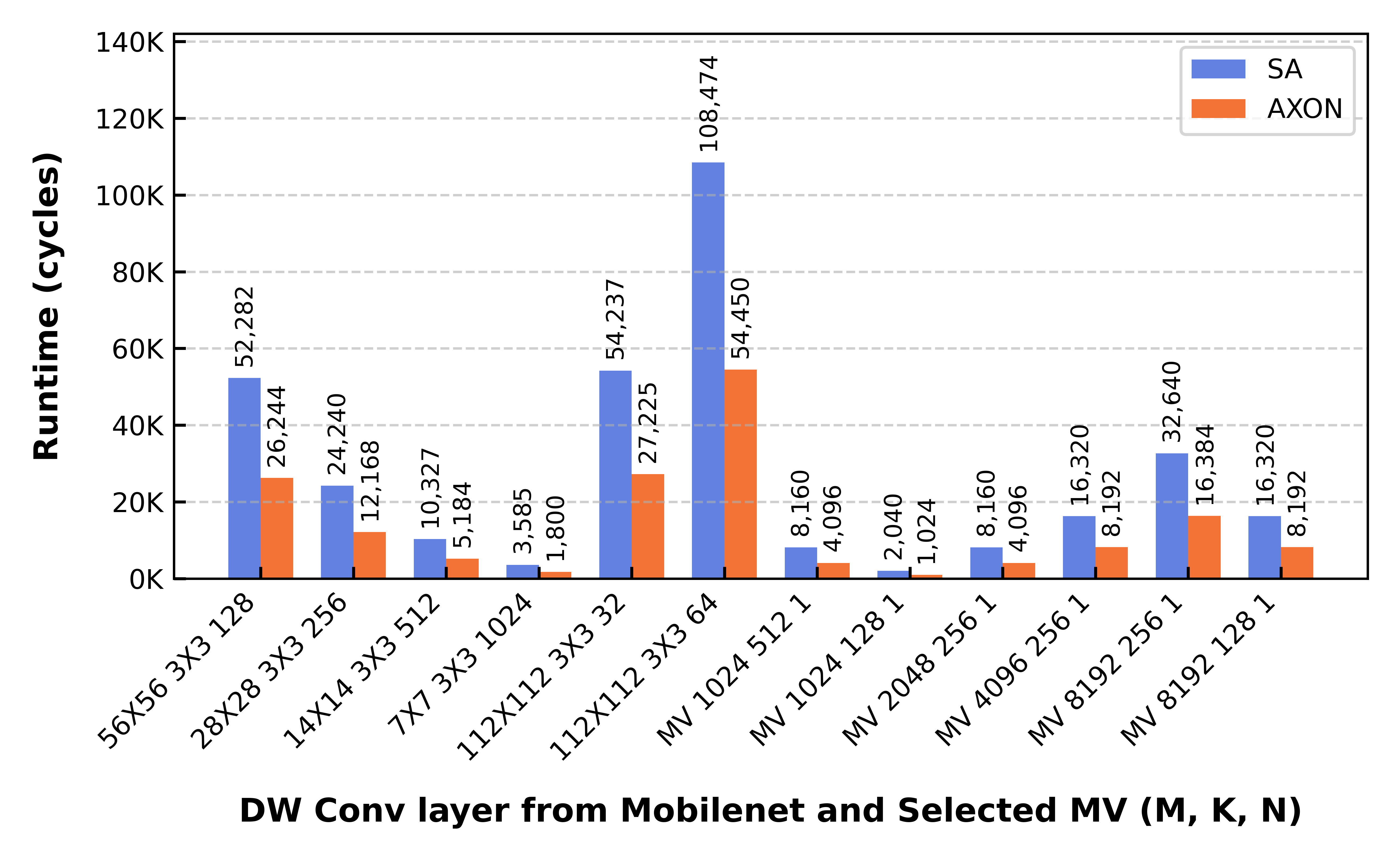}
    \vspace{-20pt}
    \caption{Runtime improvement evaluation on DW Conv and MV workloads}
    \vspace{-15pt}
    \label{fig:DW-MV perf(type2)}
    \vspace{-5pt}
\end{figure}

Besides SA, We have also compared PE utilization rate (UR) improvement over conventional SA for Axon with Configurable multidirectional systolic array (CMSA) architecture proposed by Xu et al\cite{xu2021configurable}. Fig.\ref{fig:UR_imp_comp} illustrates the comparison where UR has been calculated for array shape of $128\times128$. We observe that the utilization rate improvement varies for Axon and CMSA based on the workload shape and size. Axon outperforms CMSA by an average of $27\%$ in terms of the utilization rate improvement. It should be noted that in some workloads the improvement remains small for both cases (e.g. GPT3 matmul1, GPT3 addmm, GPT3 mhead).  This is because for those particular workloads the utilization rate is already high (average $91\%$) for conventional SA.
\vspace{-5pt}
\subsubsection{Energy and Area Comparison}
\begin{figure}[htbp]
    \centering
    \vspace{-5pt}
    \subfigure[]{\includegraphics[width=0.49\linewidth]{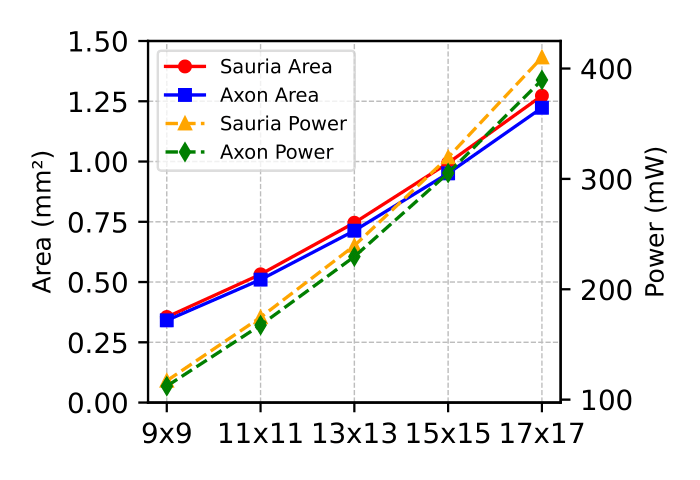}} \hfill
    \subfigure[]{\includegraphics[width=0.49\linewidth]{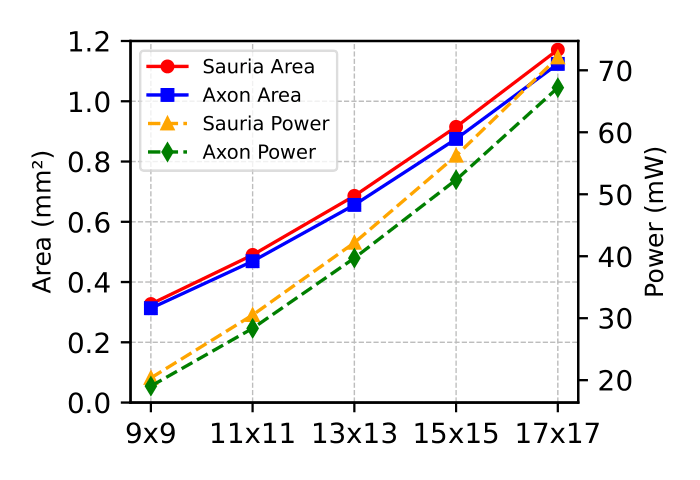}}
    \vspace{-15pt}
    \caption{Power and area comparsion with sauria SA im2col. a) 45nm node b) 7nm node}
    \label{powerAreaComp}
    \vspace{-10pt}
\end{figure}

To observe the silicon footprint and power overhead we used 45nm and 7nm advanced technology node for synthesis over different array shapes as demonstrated in Fig.\ref{powerAreaComp}a) and b) respectively. We have compared with sauria\cite{fornt2023energy} as it includes hardware im2col support like ours. We find that Axon has an average $3.93\%$ less area and $4.5\%$ less power consumption over Sauria because Axon uses 2to1 mux instead of Sauria's data feeder registers and counters for im2col support.   

\vspace{-5pt}



\section{Related Work}
Samajdar et al\cite{samajdar2022self} proposes a reconfigurable systolic array called SARA to improve mapping flexibility as well as increase data reusability through bypass wires. In MAERI\cite{kwon2018maeri} a fabric to support arbitrary dataflows has been proposed. DRACO\cite{9155063} has been proposed to optimize memory bound DNN workload from algorithm front. AI-MT\cite{9138929} is designed to maximize the accelerator's computational resources and memory bandwidth by pairing compute-intensive and memory-intensive tasks from different networks, and thus allowing for their parallel execution. COSA \cite{wang2023cosa} is developed to support hybrid data reuse by analyzing the computational characteristics of attention mechanism. However, all these works focus on supporting diverse aspect ratios that arise in DNNs. Our work is orthogonal and can be applied over them.  For hardware support of im2col, SPOTS \cite{soltaniyeh2021spots} support sparsity with on-the-fly im2col but require a large intermediate buffer. USCA \cite{liu2019usca} and SAURIA \cite{fornt2023energy} support convolution lowering by using a dedicated data feeder. In contrast, Our work offer im2col through 2to1 mux with exploiting data reusability from adjacent PE buffer.
\vspace{-5pt}
\section{Conclusion}
We present a novel data orchestration in a systolic array that outperforms conventional systolic array and state-of-the-art in terms of run time in all three dataflows (OS, IS, and WS). The Axon architecture offers faster computation (upto $2\times$ speedup) with very small power overhead thus making the design more energy efficient. The proposed architecture of Axon unified PE can leverage workloads' diverse shapes to further improve the runtime. We also propose im2col hardware support enabled by the Axon data orchestration. Axon im2col hardware support can relax the high memory bandwidth requirement by $60\%$ during convolution workloads posed by software im2col with significantly lower hardware (0.2\%) and power (1.6\%) overhead. 




\bibliographystyle{ACM-Reference-Format}
\bibliography{axon}
\end{document}